\documentclass[epj,referee]{svjour}

\usepackage{latexsym}
\usepackage{amsmath}
\usepackage{amsfonts}
\usepackage{graphicx}

\newcommand{\ave}[1]{\left\langle #1 \right\rangle}

\newcommand{\kln}{ \ln_{\kappa} }
\newcommand{\kexp}{ \exp_{\kappa} }
\newcommand{\arcsinh}{\mathop\mathrm{arcsinh}\nolimits}

\begin{document}
\title{A nonlinear drift which leads to $\kappa$-generalized distributions}

\titlerunning{A nonlinear drift which leads to $\kappa$-Gaussians}
\author{T. Wada
}                     
%
%
\institute{Department of Electrical and Electronic
Engineering,
Ibaraki University, Hitachi,~Ibaraki, 316-8511, Japan.\\
\email{wada@mx.ibaraki.ac.jp}
}
\authorrunning{Wada}

\date{Received: date / Revised version: date}
%

\abstract
{
We consider a system described by a Fokker-Planck equation
with a new type of momentum-dependent drift coefficient
which asymptotically decreases as $-1/p$ for a large momentum $p$. 
It is shown that the steady-state of this system 
is a $\kappa$-generalized Gaussian distribution, which
is a non-Gaussian distribution with a power-law tail.  
\PACS{
      {05.20.Dd}{Kinetic theory} \and
      {05.20.-y}{Classical statistical mechanics}   \and
      {05.90.+m}{Other topics in statistical physics}
     } 
} 
\maketitle
%

%


\section{Introduction}
Non-Gaussian probability distributions are frequently observed in a variety of 
systems such as physical, chemical, economical and social systems.
Known examples of non-Gaussian probability distributions are L\'evy $\alpha$-stable distributions, which can be defined by its Fourier transformation as
\begin{align}
 \mathcal{L}_{\alpha}^C(x) = \frac{1}{2 \pi} \int dk
 \exp[ikx-C\vert k \vert^{\alpha}], \quad (0<\alpha \le 2)
\end{align}
and Tsallis' $q$-generalized distributions \cite{Tsallis},
\begin{align}
  W_q(x) \propto \left[1-(1-q) \beta x^2 \right]^{\frac{1}{1-q}}, \quad (1 < q \le 3)
\label{q-Gaussian}
\end{align}
in the non-extensive statistical mechanics based on Tsallis' entropy \cite{Tsallis}.
A common key feature of the both probability distributions is the presence of an asymptotic power-law tail,
$\mathcal{L}_{\alpha}^C(x) \sim \vert x \vert^{-\alpha-1}$, and
$W_q(x) \sim x^\frac{2}{1-q}$, respectively.

There is another type of non-Gaussian distributions 
with asymptotic power-law tails, which is called a $\kappa$-generalized 
Gaussian,
\begin{align}
  W_{\kappa}(x) \propto 
\left( -\kappa \beta x^2 + \sqrt{1 + \kappa^2 \beta^2 x^4} 
       \right)^{\frac{1}{\kappa}},
 \quad (\vert \kappa \vert < 2)
  \label{k-Gaussian}
\end{align}
It has been originally studied in the context of statistical
physics by Kaniadakis \cite{k-entropy}. This $\kappa$-generalized 
Gaussian can be derived by maximizing Kaniadakis' $\kappa$-entropy 
under appropriate constraints. 
This $\kappa$-Gaussian reduces to the standard Gaussian, $\exp(-\beta x^2)$,
in the limit of $\kappa=0$.
For a large value of $x$, the $\kappa$-Gaussian
obeys a power-law as  
$W_{\kappa}(x) \sim x^{-\frac{2}{\kappa}}$.\\
The $\kappa$-generalized distributions have been shown to well explain,
for example, the energy distributions of cosmic rays \cite{k-entropy}, 
and the size distribution of personal incomes \cite{Clementi}.
In a previous work \cite{WS07,WS09}, we have studied the asymptotic behavior of 
the $\kappa$-generalized \textit{nonlinear} Fokker-Planck(FP) equation, 
which steady-state is a $\kappa$-generalized Gaussian distribution. 
Furthermore a $\kappa$-generalized Gaussian is also derived \cite{WS06}
by generalizing the log-likelihood function in Gauss' law of error,
which is an original method developed by Gauss himself to derive 
a standard Gaussian.

On the other hand,
Lutz \cite{Lutz} has recently shown an analytic prediction that 
the stationary momentum distributions of trapped atoms in an optical lattice 
are, in fact, Tsallis' $q$-generalized Gaussian \eqref{q-Gaussian}.
Later, Gaeta \cite{Gaeta} showed its invariance under the asymptotic Lie symmetries.
The prediction was experimentally verified by a London team \cite{Douglas}.
This anomalous transport is described by a \textit{linear} FP equation 
with a nonlinear drift coefficient,
\begin{align}
  K^{\rm ol}(p) = - \frac{\alpha p}{1+ \left( \frac{p}{p_c} \right)^2 },
  \label{K}
\end{align}
which represents a capture force with damping coefficient $\alpha$,
and this force acts only on slow particles whose momentum is smaller than
the capture momentum $p_c$.

A characteristic feature of this nonlinear drift is that: 
for a small momentum $\vert p \vert <p_c$, the drift is approximately
linear $K^{\rm ol}(p) \sim -p$, i.e., it reduces to a familiar Ornstein-Uhlenbeck process;
whereas for a large momentum $\vert p \vert >p_c$, it asymptotically 
decreases as $K^{\rm ol}(p) \sim -1/p$.\\
%
In contrast to most systems with power-law distributions which are often
described by nonlinear kinetic equations \cite{Frank}, the above process
is described by an ordinary linear FP equation. Consequently standard
methods can be applied to the analysis of the problem.

It is worth stressing that the Lutz analysis is not restricted to anomalous transport
in an optical lattice, but can be applied to a wide class of systems
described by a FP equation with a drift coefficient decaying
asymptotically as $-1/p$.

In this contribution, we propose another momentum-dependent drift coefficient
$K(p)$ given by equation \eqref{kDrift}, which also asymptotically decreases
as $-1/p$ for a large momentum $\vert p \vert > p_c$.
We consider the process described by the linear FP equation with
this drift coefficient $K(p)$.
Next section provides a brief review of $\kappa$-generalized thermostatistics
and some properties of $\kappa$-generalized Gaussian.
In section three we consider an ordinary linear FP equation with the
proposed momentum-dependent drift coefficient $K(p)$ and a constant diffusion coefficient $D$.
It is shown that the steady-state of the FP equation
with this nonlinear drift coefficient $K(p)$ 
is a $\kappa$-generalized Gaussian.
The deformed parameter $\kappa$ can be expressed in terms
of the microscopic parameters.
In section four the asymptotic behavior of the FP equation is studied.
It is shown that the non-increase of the Lyapunov functional associated
with the FP equation. Then we numerically analyze the time evolutions
of numerical solutions against different initial probability distributions,
and show the asymptotic convergence of the numerical solutions to 
$\kappa$-Gaussian.
In section five we discuss the relation between $\beta$ and
the average energy in the parameter region that the mean-kinetic
energy diverges.
The final section is summary.

\section{$\kappa$-generalized thermostatistics}
We first give the brief review of the generalized thermostatistics based on 
$\kappa$-entropy defined as
\begin{align}
  S_{\kappa} &\equiv -k_{\rm B} \int_{-\infty}^{\infty} dp \; w(p) \kln w(p),
\end{align}
for a probability distribution $w(p)$ of the momentum $p$.
Here $k_{\rm B}$ denotes the Boltzmann constant, and  $\kln(x)$ is the $\kappa$-logarithmic function defined by
\begin{align}    
  \kln(x) \equiv \frac{x^{\kappa} - x^{-\kappa}}{2 \kappa}.
\end{align}
The $\kappa$-entropy $S_{\kappa}$ is a real-parameter ($\kappa$) extension of 
the standard Boltzmann-Gibbs-Shannon (BGS) entropy.
The inverse function of $\kln(x)$ is expressed as
\begin{align}
\kexp(x) &\equiv \exp\left[\frac{1}{\kappa} \arcsinh(\kappa x) \right]
\nonumber \\
&= \left( \kappa x + \sqrt{1 + \kappa^2 x^2} 
       \right)^{\frac{1}{\kappa}},
 \label{kexp}
\end{align}
and called $\kappa$-exponential function.
For a small value of $x$, the $\kappa$-exponential function
is well approximated with $\exp(x)$, whereas a large positive value of $x$,
it asymptotically obeys a power-law $\kexp(x) \sim x^{1/\kappa}$. 
In the limit of $\kappa=0$ both $\kln(x)$ and $\kexp(x)$ reduce to 
the standard logarithmic and exponential functions, respectively.
Accordingly the $S_{\kappa}$ reduces to the BGS entropy.

Maximizing the $\kappa$-entropy $S_{\kappa}$ under the constraints of 
the mean kinetic energy and the normalization of probability distribution $w(p)$,
\begin{align}
   \frac{\delta}{\delta w} \Big( 
     S_{\kappa}[w]-\beta \int_{-\infty}^{\infty} dp \, \frac{p^2}{2} w(p)
-\gamma \int_{-\infty}^{\infty} dp  w(p) \Big) = 0,
\end{align}
leads to a so-called $\kappa$-generalized Gaussian,
\begin{align}
 w^{\rm ME}(p) = \alpha \,\kexp
  \left[- \frac{1}{\lambda} \big(\gamma + \beta \, \frac{p^2}{2} \big)\right].
\end{align}
Here $\gamma$ is a constant for the normalization, and depends
on $\beta$, which controls the variance of $w^{\rm ME}(p)$.
The parameter $\alpha$ and $\lambda$ are $\kappa$-dependent constants,
which are given by 
\begin{align}
 \alpha &= \left(\frac{1-\kappa}{1+\kappa}\right)^{\frac{1}{2\kappa}}, \quad
 \lambda = \sqrt{1-\kappa^2},
\end{align}
respectively.

The $\kappa$-generalization of free-energy was studied in \cite{SW06}
and given by
\begin{align}
  F_{\kappa} &\equiv -\left( \frac{I_{\kappa} +\gamma}{\beta} \right),
\end{align}
where
\begin{align}
 I_{\kappa} &\equiv \int_{-\infty}^{\infty} dp \, \frac{1}{2} \Big[
  \left(w^{\rm ME}(p)\right)^{1+\kappa}+ \left(w^{\rm ME}(p) \right)^{1-\kappa}  \Big].
\end{align}
The $\kappa$-generalized free-energy $F_{\kappa}$ satisfies the Legendre transformation structures, 
\begin{align}
F_{\kappa} = U - \frac{1}{\beta} \, S_{\kappa}, \quad
   \frac{d}{d \beta} \, \Big( \beta F_{\kappa} \Big) = U,
\label{Fk}
\end{align}
where
\begin{align}
  U = \int_{-\infty}^{\infty} dp \, \frac{p^2}{2} w^{\rm ME}(p).
\end{align}

\section{Proposed nonlinear drift coefficient}
Let us consider the linear FP equation
\begin{align}
  \frac{\partial}{\partial t} w(p, t) =
  -\frac{\partial}{\partial p} \Big( K(p) \, w(p, t) \Big)  
  + D \frac{\partial^2}{\partial p^2} w(p,t),
\label{kFPE}
\end{align}
with  a constant diffusion coefficient $D$
and the momentum-dependent drift coefficient,
\begin{align}
  K(p) = -\frac{\alpha p}{\sqrt{1+\left(\frac{p}{p_c}\right)^4}},
\label{kDrift}
\end{align}
where $\alpha$ is a damping coefficient and $p_c$ denotes a capture
momentum.
Note that this proposed drift coefficient $K(p)$ also 
behaves as $-p$ for a small momentum $\vert p \vert <p_c$,
and asymptotically decreases
as $-1/p$ for a large momentum $\vert p \vert > p_c$
as same as $K^{\rm ol}(p)$ in anomalous diffusions in optical lattice \cite{Lutz}.
We introduce the associated potential,
\begin{align}
  V(p) = \frac{p_c^2}{2} \arcsinh \left( \frac{p^2}{p_c^2} \right),
\label{k-potential}
\end{align}
which is related with $K(p)$ by
\begin{align}
  K(p) = -\alpha \frac{d}{dp} \, V(p).
  \label{K-V}
\end{align}

\subsection{Steady-state}
Next we show the steady-state $w_s(p)$ of the FP equation with the nonlinear
drift \eqref{kDrift} is a $\kappa$-Gaussian.
To this, the  steady-state condition $\frac{\partial}{\partial t} w_s(p) =0$
in Eq. \eqref{kFPE} leads to 
\begin{align}
  \frac{d}{d p} \ln w_s(p) &=
\frac{K(p)}{D} = - \frac{d}{dp} \frac{\alpha V(p)}{D}.
\end{align}
In the last step, we used the relation \eqref{K-V}.
Substituting equation \eqref{k-potential} and after integration we have
\begin{align}
  \ln w_s(p) = -\frac{\alpha p_c^2}{2 D} \, \arcsinh\left(\frac{p^2}{p_c^2}\right) + \textrm{const. },
\end{align}
then the steady-state becomes
\begin{align}
w_s(p) &\propto \exp\left[-\frac{\alpha p_c^2}{2 D} 
 \arcsinh \left(\frac{p^2}{p_c^2}\ \right)   \right].
\end{align}
By using the definition \eqref{kexp} of $\kappa$-exponential function, and
introducing the two parameters as
\begin{align}
 \kappa = \frac{2 D}{\alpha p_c^2},\quad
 \beta = \frac{\alpha}{D},
\label{para}
\end{align}
we can express the steady-state as  
\begin{align}
  w_s(p) = \frac{1}{Z_{\kappa}} \kexp \left[-\beta \frac{p^2}{2} \right],
\end{align}
where $Z_{\kappa}$ is the normalization factor.
We thus found that the steady-state of the FP equation with the nonlinear
drift coefficient $K(p)$ is nothing but a $\kappa$-generalized Gaussian.

Remarkably, the parameter $\kappa$ can be expressed in terms
of the microscopic parameters, i.e., $\alpha, D, p_c$ in the FP equation.
This fact allows us to give a physical interpretation of 
the $\kappa$-generalized distribution, as similar as $q$-generalized distribution in Lutz' analysis \cite{Lutz}.
For example, in the limit of $p_c \to \infty$, the drift coefficient $K(p)$
reduces to $-p$ of the standard Ornstein-Uhlenbeck process, and
the deformed parameter $\kappa$ of equation \eqref{para} reduces to $0$.
This corresponds to the standard case in which the steady-state $w_s(p)$
is a standard Gaussian.  \\
Note also that the parameter $\beta$ is expressed as the ratio of
the friction coefficient $\alpha$ to the diffusion coefficient $D$,
in analogy with  the fluctuation-dissipation relation.
We emphasize that the parameter $\beta$ is not equal to an inverse temperature, 
because $w_s(p)$ is not an equilibrium state but a steady-state, for which
temperature is not well defined. 

\section{Asymptotic behavior}
We here study the asymptotic solutions of the FP equation 
with the nonlinear drift coefficient $K(p)$.\\
In a previous work \cite{WS07,WS09} we have studied the nonlinear FP equation
associated the $\kappa$-generalized entropy, and shown the existence
of the associated Lyapunov functional, which characterizes a long
time behavior of the process described by the FP equation.
Similarly, the Lyapunov functional, 
\begin{align}
  {\mathcal L}(t) \equiv U[w] - \frac{D}{\alpha} \, S[w],
  \label{L}
\end{align}
is monotonically non-increasing, i.e., $ \frac{d} {dt} {\mathcal L}_{\kappa}(t) \le 0$, for any time evolution of $w(x,t)$ according to the linear FP
equation \eqref{kFPE}.
In equation \eqref{L}, 
\begin{align}
  S[w] = - k_{\rm B} \int dp \; w(p, t) \ln w(p, t),
\end{align}
is BGS entropy, and
\begin{align}
  U[w] \equiv \int dp \; V(p) \, w(p, t).
\end{align}
is the ensemble average of the potential $V(p)$.\\
The proof of the non-increase of ${\mathcal L}(t)$ is as follows:
\begin{align}
   \frac{d {\mathcal L}(t)}{d t}  =& \int_{-\infty}^{\infty} dp \; 
       \frac{\partial }{\partial w} \left[ V(p)\, w
         + \frac{D}{\alpha} w \ln w \right] \;  
  \frac{\partial w(p, t)}{\partial t} \nonumber \\
    =& \int_{-\infty}^{\infty} dp \; \left[ V(p)
         + \frac{D}{\alpha}( \ln w + 1 ) 
     \right]
\nonumber \\
  &\qquad \times \frac{\partial}{\partial p}\,\left[-K(p) w +
   D \frac{\partial}{\partial p}\, w \right]
\nonumber \\
 = & -\int_{-\infty}^{\infty} dp \, \frac{w}{\alpha}
     \left\{ -K(p) + D \, \frac{\partial}{\partial p} \ln w \right\}^2
 \le 0.
\end{align}
From the fist line to the second line we used the FP equation \eqref{kFPE},
and in the last step we used the integration by part.\\
Thus ${\mathcal L}(t)$ is non-increasing and
consequently ${\mathcal L}(t)$ is minimized for 
the steady-state $w_s(p)$ as
\begin{align}
 \min {\mathcal L}(t) = \lim_{t \to \infty} {\mathcal L}(t) 
= U[w_s(p)] - \frac{D}{\alpha} \, S[w_s(p)].
 \label{freeE}
\end{align}
Note that the last expression is the free-energy associated
with the steady-state $w_s(p)$ of the linear FP equation.
In contrast to the $\kappa$-generalized free-energy \eqref{Fk}
associated with the nonlinear FP equation \cite{WS07,WS09},
$S$ is the standard BGS entropy and
$U$ is the ensemble average of the nonlinear potential $V(p)$
in the relation \eqref{freeE}. 

\subsection{Asymptotic convergence to $\kappa$-Gaussian}
In order to study a long time behavior of the FP equation
with the nonlinear drift coefficient $K(p)$,
we performed numerical simulations against different initial probability
distributions.
We used a variant of the numerical method originally developed 
by Gosse and Toscani \cite{GT06} for the Cauchy problem an the
evolution equation.
For the details of the numerical scheme, please refer to \cite{GT06}.

A time-evolution of the numerical solution $w(p,t)$ of the
FP equation with $K(p)$ is shown in figure 1.
\begin{figure}
\begin{center}
 \includegraphics[scale=0.9]{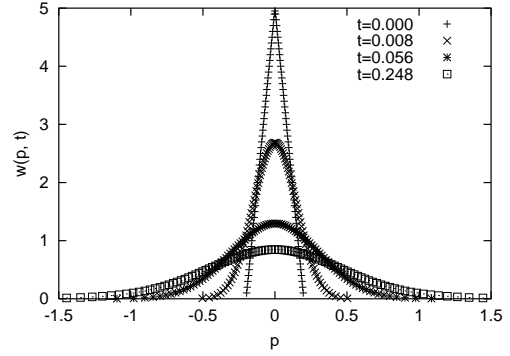}
\caption{
A typical time-evolution of $w(p,t)$ from an initial probability density 
with triangle shape.
The microscopic parameters are set to $p_c=1, D=1, \alpha=4$,
so that $\kappa=0.5, \beta=4$.
}
\end{center}
\label{evo}
\end{figure}
Note that the numerical solution $w(p,t)$ seems to be asymptotically
approaching to the $\kappa$-Gaussian.
In order to confirm this property, we 
fitted the numerical solution $w(p,t)$ at each time $t$ with 
the $\kappa$-Gaussian
\begin{align}
   a(t) \kexp \left[- b(t) p^2 \right],
   \label{k-Gau}
\end{align}
where $a(t)$ and $b(t)$ are fitting parameters.
Then the time evolution of the function defined by
\begin{align}
   \eta(p,t) \equiv 
 \ln \left( \frac{w(p, t)}{a(t) \kexp \left[- b(t) p^2 \right]} \right),
  \label{ratio}
\end{align}
are studied. If a numerical solution $w(p,t)$ is perfectly fitted
with equation \eqref{k-Gau}, the function $\eta(p,t)$ becomes
identically zero. In figure 2 the time evolutions
of $\eta(p,t)$ and $w(p,t)$ against an initial probability distribution 
$w(p,0)$ with a triangle shape are plotted. It is obvious from this figure that the function $\eta(p,t)$
is decreasing to zero as time evolves. This fact shows that 
the numerical solutions are approaching to the $\kappa$-Gaussian.
 
\begin{figure}
\begin{center}
\includegraphics[width=0.38\textwidth]{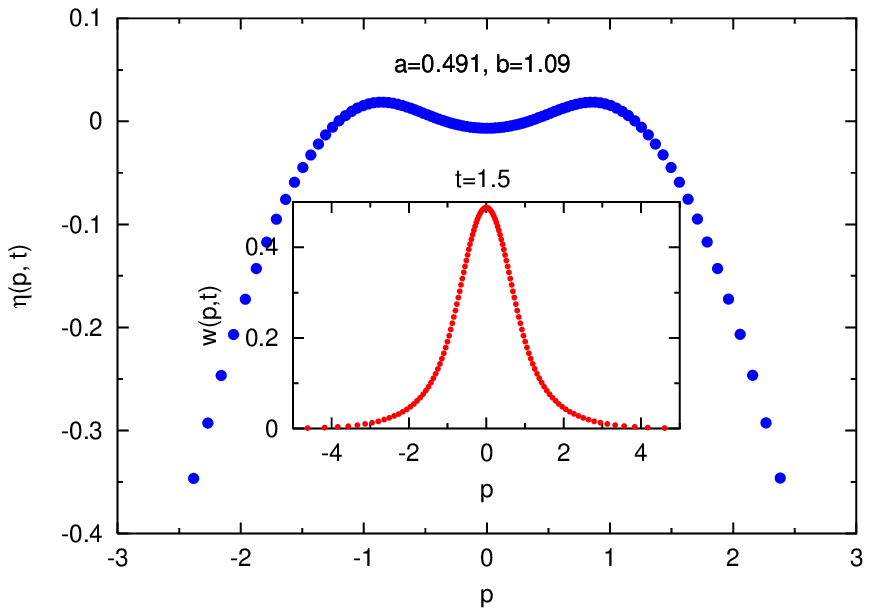}
\includegraphics[width=0.38\textwidth]{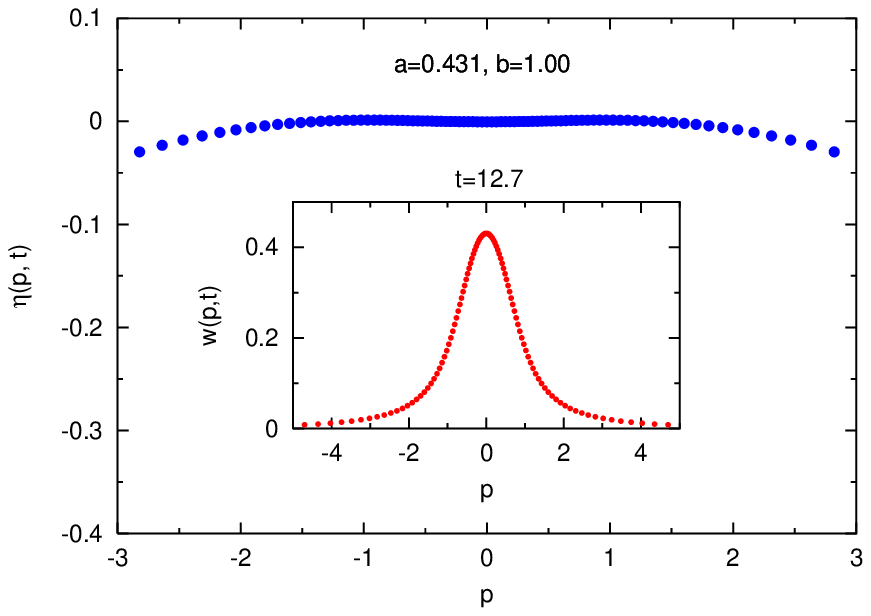}
\includegraphics[width=0.38\textwidth]{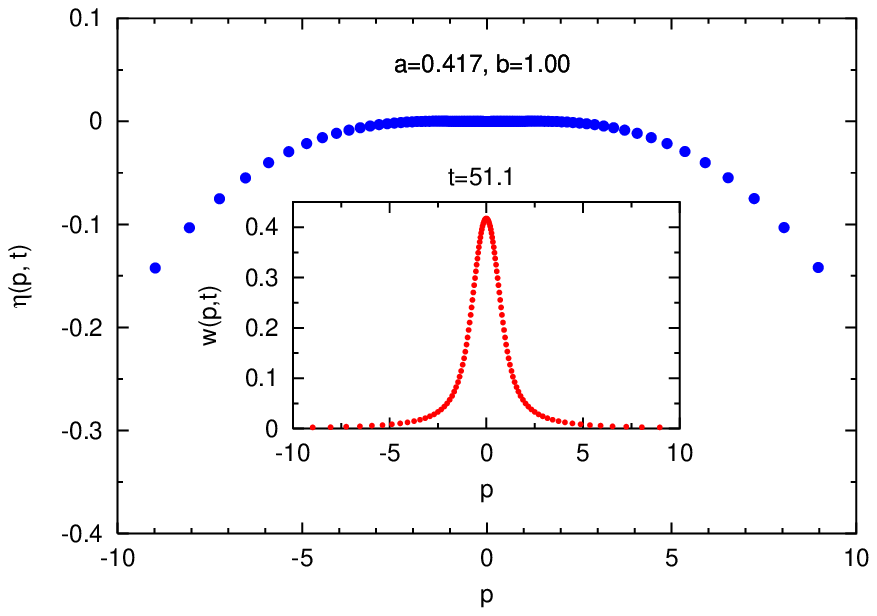}
\includegraphics[width=0.38\textwidth]{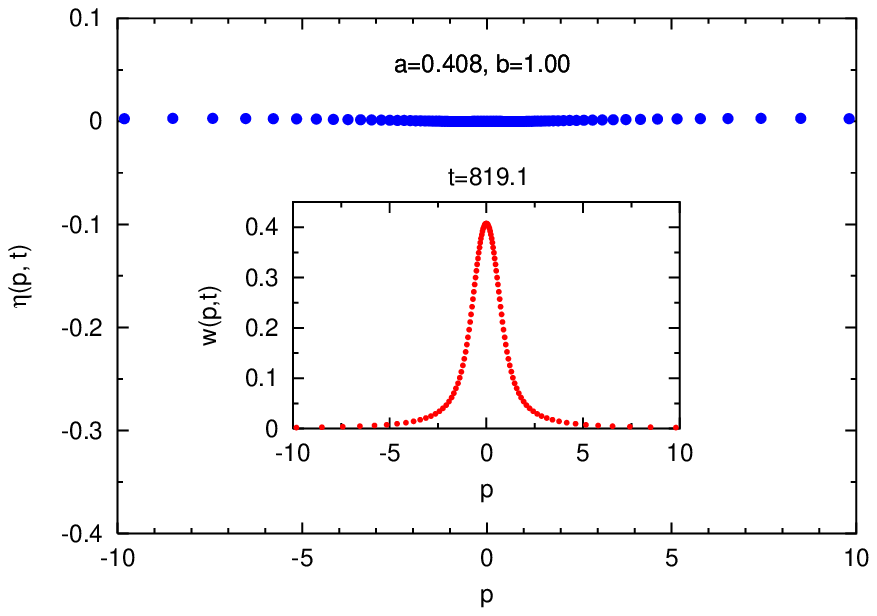}
\caption{Long time behavior of the quantity $\eta(p,t)$ given in 
equation \eqref{ratio}
and the numerical solutions $w(p,t)$ against
an initial probability distribution with triangle shape. 
The microscopic parameters are set to $p_c=1, D=1, \alpha=2$, 
hence $\kappa=1, \beta=2$.
The number of calculated points are $101$.
The best fitted parameters $a(t)$ and $b(t)$ in \eqref{k-Gau} are indicated
in each plot.
Inset figure shows $w(p,t)$ at each time $t$.}
\end{center}
\label{eta-evolution}
\end{figure}

\section{The relation between $\beta$ and the average energy}

The $\kappa$-Gaussian of the steady-state $w_s(p)$ is not
normalizable for the parameter region of $ 2 \le \vert \kappa \vert$,
or equivalently, $\alpha p_c^2 \le D$. As similarly as in the momentum
distributions in an optical lattice, the physical meaning of this
is that compared with the random momentum fluctuations $D$, the capture force 
$\alpha p_c^2$ is too weak to keep the particle around the bottom ($p=0$)
of the potential $V(p)$. In other words, the potential is too shallow
to capture a particle with the large random momentum fluctuations. 

Next let us turn our focus on the parameter region of 
$2/3 < \vert \kappa \vert < 2 (D < \alpha p_c^2 < 3 D)$, 
in which the second moment
\begin{align}
\ave{p^2} \equiv \int_{-\infty}^{\infty} dp \, p^2 \, w_s(p), 
\end{align}
of the $\kappa$-Gaussian becomes infinite.
Consequently the mean kinetic energy, $\ave{p^2}/2m$, diverges,
and it is the hallmark of an anomalous diffusion.
Lutz \cite{Lutz} showed that an explicit correspondence
between ergodicity breaking in a system described by power-law tail distributions and the divergences of the moments of these distributions, i.e.,
Ensemble average and time average of the dynamical variable $p^n$
are not equal to each other in the infinite-time limit,
when the $2n$-th moment of the stationary momentum distribution 
for a system described by power-law tail distributions diverges.
His analysis is also valid for the present study because both momentum
dependent drift coefficient $K^{\rm ol}(p)$ and $K(p)$ have the same asymptotic
behavior $\sim -1/p$, and consequently both steady-states are non-Gaussian
distributions with power-law tails.   

Whereas the mean kinetic energy diverges in this way,
in the same region, let us consider the following average energy
\begin{align}
   \ave{ p \frac{d}{dp} V(p)} &\equiv
  \int_{-\infty}^{\infty} \frac{p^2}{\sqrt{1+\left(\frac{p}{p_c}\right)^4}}
 \, w_s(p).
\label{ave}
\end{align}
Since
\begin{align}
 - \frac{1}{\beta} \frac{d}{dp} \, \kexp\left(-\beta \frac{p^2}{2} \right)
= \frac{p\kexp\left(-\beta \frac{p^2}{2} \right)}{\sqrt{1+\left(\frac{p}{p_c}\right)^4}},
\end{align}
integrating by part, the r.h.s. of equation \eqref{ave} becomes
\begin{align}
-\frac{p}{\beta}  \kexp \left(-\beta \frac{p^2}{2} \right) \Big\vert_{-\infty}^{\infty}+
\frac{1}{\beta} \int_{-\infty}^{\infty} dp \, w_s(p)
\end{align}
In the region $2/3 < \vert \kappa \vert < 2$, the first term become zero
and the $w_s(p)$ is normalizable, thus we finally obtain
\begin{align}
   \ave{ p \frac{d}{dp} V(p)} = \frac{1}{\beta}.
\label{this}
\end{align}
This relation remind us \textit{a general equipartition principle} \cite{Tolman},
\begin{align}
   \ave{ p \frac{\partial}{\partial p} E} = k_{\rm B} T,
\label{GER}
\end{align}
where $E$ is the energy of a system in thermal equilibrium
with the temperature $T$. However, as pointed out before,
$\beta$ is not an inverse temperature since the steady-state $w_s(p)$
is, in general, a non-equilibrium state, in which the temperature is not
well defined.


\section{Summary}
We have proposed a momentum-dependent drift coefficient which
asymptotically decreases as $-1/p$  for a large momentum $\vert p \vert > p_c$.
We have studied a system described by the FP equation with this drift
coefficient, and found that 
the steady-state is a $\kappa$-generalized probability distribution.
We performed the several numerical simulations in order
to study asymptotic behaviors of the numerical solutions against the different
initial probability distributions, and found that these numerical solutions 
asymptotically approach to the $\kappa$-Gaussian functions.

\section*{Acknowledgement}
This research was partially supported by the Ministry of Education,
Science, Sports and Culture (MEXT), Japan, Grant-in-Aid for Scientific
Research (C), 19540391, 2008.

\end{document}